\documentclass[epj]{svjour}
\usepackage{latexsym}
\usepackage{graphicx}
\usepackage[usenames,dvipsnames]{xcolor} 
\usepackage{amsmath,amssymb}

\begin{document}
\title{Overspinning Kerr-MOG black holes by test fields and the third law of black hole dynamics}
\author{Koray D\"{u}zta\c{s}%\inst{1} %\and Second author\inst{2}% etc
}                     % Do not remove

\institute{Department of Natural and Mathematical Sciences,
\"{O}zye\u{g}in University, 34794 \.{I}stanbul Turkey }

\date{Received: date / Revised version: date}
\abstract{We evaluate the validity of the weak form of the cosmic censorship conjecture and the third law of black hole dynamics for Kerr-MOG black holes interacting with scalar test fields. Ignoring backreaction effects, we first show that both extremal and nearly extremal Kerr-MOG black holes can be overspun into naked singularities by scalar test fields with a frequency slightly above the superradiance limit. In addition,  nearly extremal Kerr-Mog black holes can be continuously driven to extremality by test fields. Next, we employ backreaction effects based on the argument that the angular velocity of the event horizon increases before the absorption of the test field. Incorporating the backreaction effects, we derive that the weak form of the cosmic censorship and the third law are both valid for Kerr-Mog black holes with a modification parameter $\alpha \lesssim 0.03$, which includes the Kerr case with $\alpha=0$. 
%Insert your abstract here.
%
 % end of PACS codes
} %end of abstract
\maketitle
\section{Introduction}
The singularity theorems developed by Penrose and Hawking imply that the gravitational collapse of a body leads --inevitably-- to the formation of singularities~\cite{singtheo}. The presence of these singularities precludes the definition of a well-defined initial value problem and thereby ruins the smooth, deterministic structure of space-times in general relativity. The fact that the formation of singularities cannot be avoided led Penrose to propose the cosmic censorship conjecture, which --in its weak weak form (Wccc)-- asserts that the gravitational collapse of a body always ends up in a black hole rather than a naked singularity~\cite{ccc}. The singularities should be hidden behind the event horizons of black holes which disable their causal contact with distant observers. This way, the observers at asymptotically flat spatial infinity do not encounter any effects propagating out of the singularity, and the smooth structure of space-times is preserved, at least locally. 

As a concrete proof of the cosmic censorship conjecture has been elusive, it has become customary to attack the closely related --though not identical-- problem of the stability of event horizons. In these problems one perturbs extremal or nearly extremal black holes with test particles and fields, and checks if the perturbations can lead to the destruction of event horizons which would imply that the singularities become naked. The first thought experiment in this vein was constructed by Wald \cite{wald74}. There it was shown that test particles cannot overcharge or overspin an extremal Kerr-Newman black hole into a naked singularity. Following Wald many similar tests of Wccc were applied to black holes in electro-vacuum spacetimes involving test particles \cite{hu,Jacobson-Sot,f1,saa,gao,siahaan,magne,dilat,higher,v1}, and fields \cite{semiz,emccc,overspin,duztas,toth,natario,duztas2,mode,w2,taub-nut,kerrsen}. The possibility to violate Wccc by the quantum tunnelling of test particles was discussed in \cite{q1,q2,q3,q4,q5,q6,q7}. The stability of event horizons in the asymptotically anti-de Sitter case was also evaluated by perturbing the black holes with test particles and fields~\cite{btz,gwak1,gwak2,gwak3,gwak4,chen}.

The evolution of singularities indicate the failure of general relativity at short length scales where quantum effects are expected to dominate. In addition, the fact that one needs to invoke the presence of dark components at large length scales motivated the quest for modified theories of gravity. One of the promising candidates to fill this gap is the Scalar-Tensor-Vector Gravity theory developed by Moffat~\cite{stvg}. This dark matter emulating theory of modified gravity has proved compatible with current observations regarding the rotation curves of galaxies and the dynamics of galactic clusters \cite{cluster1,cluster2,cluster3,cluster4}. It also predicts the existence of gravitational waves which lends credence to its validity as an alternative theory of gravity \cite{gwave1,gwave2}. 

The scalar-tensor-vector theory of modified gravity has a stationary and axi-symmetric black hole solution which is known as the Kerr-MOG black hole \cite{kerrmog}. Kerr-MOG black holes are characterised by the mass parameter $M$, angular momentum $J=Ma$ and the dimensionless parameter $\alpha$ which determines the modification from the Kerr solution. The thermodynamics of Kerr-MOG black holes, their observable shadows, and the quasi-normal modes have been studied \cite{thermo,shadow,quasi}. Recently, it was also shown that energy can be extracted from Kerr-MOG black holes by a Penrose process \cite{pradhan}.

The validity of Wccc was tested for Kerr-MOG black holes in the process of the absorption of a point particle by Liang, Wei, and Liu \cite{liang}. It was found that --though the extremal black holes cannot-- nearly-extremal black holes can be destroyed by point particles. However, the authors argued that the event horizon will be restored when one considers the effect of the adiabatic process. Another intriguing problem at this stage is to test the validity of Wccc in the case of test fields scattering off Kerr-MOG black holes. In this work we evaluate the stability of the event horizons of Kerr-MOG black holes as they are perturbed by test scalar fields. We consider the cases of both extremal and near-extremal black holes. Our analysis exploits the fact that superradiance occurs when scalar fields scatter off Kerr-MOG black holes, which was recently derived by Wondrak, Nicolini, and Moffat \cite{wondrak}. We also evaluate the validity of the third law of black hole dynamics which states that a nearly extremal black hole cannot be driven to extremality by any continuous process. 

\section{Kerr-MOG black holes, scalar fields, Wccc}
\label{overspin}
In Boyer-Lindquist coordinates, the background geometry of the Kerr-MOG space-time is described by the metric
\begin{eqnarray}
ds^2 &=& -\frac{\Delta}{\rho^2} [dt-a\sin^2 \theta d\phi ]^2 +\rho^2 \left[ \frac{dr^2}{\Delta} + d\theta^2 \right] \nonumber \\
&+&\frac{\sin^2 \theta}{\rho^2} \left[ (r^2+a^2)d\phi -adt \right]^2
\end{eqnarray}
where
\begin{eqnarray}
\rho^2 &=& r^2 + a^2 \cos^2 \theta \nonumber \\
\Delta &=& r^2-2G_{\rm{N}}(1+\alpha)Mr + a^2 + G^2_{\rm{N}}\alpha (1+\alpha) M^2
\end{eqnarray}
The MOG parameter $\alpha$ is a dimensionless measure of the difference between the Newtonian gravitational constant $G_{\rm{N}}$ and the additional gravitational constant $G$
\begin{equation}
\alpha=\frac{G-G_{\rm{N}}}{G_{\rm{N}}}
\end{equation}
The ADM mass and the angular momentum of the Kerr-MOG black hole are given by \cite{adm}
\begin{equation}
\mathcal{M}=(1+ \alpha)M \quad; J=\mathcal{M} a
\end{equation}
The function $\Delta$ can be re-written in terms of the ADM mass
\begin{equation}
\Delta=r^2 -2\mathcal{M}r+a^2+\frac{\alpha}{1+\alpha}\mathcal{M}^2 \label{delta}
\end{equation}
where we have set $G_{\rm{N}}=1$ without loss of generality. The spatial locations of the horizons are the roots of $\Delta$

\begin{equation}
r_{\pm}=\mathcal{M} \pm \sqrt{\frac{\mathcal{M}^2}{1+\alpha}-a^2} \label{rplus}
\end{equation}
Notice that the parameters of the Kerr-MOG space-time represent a black hole surrounded by an event horizon provided that
\begin{equation}
\mathcal{M}^2 \geq (1+\alpha) a^2 \label{criterion}
\end{equation}
where the equality corresponds to the case of an extremal black hole. In this work, we start with a Kerr-MOG black hole satisfying the main criterion (\ref{criterion}), and perturb the space-time with a scalar field that is incident on the black hole from infinity. In this type of gedanken experiments it is a crucial
assumption that the interaction of the black hole with the
test scalar field does not alter the structure of the background
geometry, but leads to modifications in the ADM mass and
angular momentum parameters. At the end of the interaction the field decays away, leaving behind a space-time with perturbed parameters. If the final parameters of the space-time does not satisfy the inequality (\ref{criterion}), one can conclude that the event horizon has been destroyed in the interaction of the scalar field with the black hole; i.e. Wccc is violated. 

The scattering of test scalar fields by Kerr-MOG black holes has recently been studied by Wondrak, Nicolini, and Moffat \cite{wondrak}. Analogous to the Kerr case, a neutral wave can be separated into variables in the form
\begin{equation}
\Psi (t,r,\theta, \phi)=R(r)S(\theta)e^{im\phi}e^{-i\omega t}
\end{equation}
The contribution of the scattering wave to the mass and angular momentum parameters of the space-time are related by
\begin{equation}
\frac{\delta  \mathcal{M}}{\delta J}=\frac{\omega}{m} \label{beken}
\end{equation}
Superradiance occurs for scalar fields scattering off Kerr-MOG black holes as one would naively expect from Kerr analogy. If the frequency of the incoming wave is below the superradiance limit, the wave is reflected back with a larger amplitude, i.e. there is no net absorption of the wave by the black hole. The superradiance limit $\omega_{\rm{sl}}$ for Kerr-MOG black holes is also derived in \cite{wondrak}
\begin{equation}
\omega_{\rm{sl}}=m\Omega=\frac{ma}{r_+^2 +a^2} \label{superrad}
\end{equation}
where $\Omega$ is the angular velocity of the black hole and $r_+$ is the spatial location of the event horizon.
\subsection{Overspinning extremal Kerr-MOG black holes}
By definition, an extremal Kerr-MOG black hole satisfies
\begin{equation}
\delta_{\rm{in}}=\mathcal{M}^2-J\sqrt{1+\alpha}=0 \label{ex}
\end{equation}
where we have defined $\delta_{\rm{in}}$. We perturb the extremal black hole with a scalar field to check if it is possible to overspin the black hole into a naked singularity. The contribution of the incoming wave to the energy and angular momentum parameters of the black hole are related by (\ref{beken}). A necessary condition for overspinning to occur is that  one should be able to adjust the parameters of the incoming wave such that $\delta_{\rm{fin}}<0$ at the end of the interaction. To be more precise, we demand that
\begin{eqnarray}
\delta_{\rm{fin}}&=&\mathcal{M}_{\rm{fin}}^2-J_{\rm{fin}} \sqrt{1+\alpha} \nonumber  \\
&=&(\mathcal{M}+\delta E)^2 -(J+\delta J)\sqrt{1+\alpha}<0 \label{condiex}
\end{eqnarray}
By substituting $\delta J=(m/\omega) \delta E$, and using (\ref{ex}), the condition (\ref{condiex}) can be simplified in the form
\begin{equation}
\delta E +2\mathcal{M}<\frac{m}{\omega}\sqrt{1+\alpha}
\end{equation}
We choose $\delta E=\mathcal{M} \epsilon$ for the incoming field wit $\epsilon \ll 1$, so that the test field approximation is justified. With this choice we can derive the maximum frequency of an incoming wave, which can be used to overspin an extremal Kerr-MOG black hole
\begin{equation}
\omega <\omega_{\rm{max}}=\frac{m\sqrt{1+\alpha}}{\mathcal{M}(2+\epsilon)} \label{condiex1}
\end{equation} 
If the frequency of a scalar field is below the maximum value determined in (\ref{condiex1}), the scalar field can overspin an extremal Kerr-MOG black hole into  a naked singularity. However, this condition is  not sufficient for overspinning to occur. For that purpose one should also demand that the incoming wave is absorbed by the black hole; i.e. the frequency of the wave is larger than the superradiance limit. These two conditions should be simultaneously satisfied for overspinning to occur. The superradiance limit for extremal black holes can be derived by substituting $r_+=M$ and and $a=M/(1+\alpha)$ in (\ref{superrad})
\begin{equation}
\omega_{\rm{sl}}=\frac{m\sqrt{1+\alpha}}{\mathcal{M}(2+\alpha)} \label{superradex}
\end{equation}
For overspinning to occur $\omega_{\rm{max}}$ should be larger than the superradiant limit $\omega_{\rm{sl}}$, so that the frequencies in the range $(\omega_{\rm{sl}},\omega_{\rm{max}})$ can be used to overspin an extremal Kerr-MOG black hole. It is manifest in equations (\ref{condiex1}) and (\ref{superradex}) that $\omega_{\rm{max}}$ will larger than  $\omega_{\rm{sl}}$, if $\alpha>\epsilon$. The extremal Kerr-MOG black holes can be overspun into naked singularities by scalar test fields provided that the deformation parameter $\alpha$ is larger than the small parameter $\epsilon$.
\subsection{Overspinning nearly-extremal Kerr-MOG black holes}
\begin{center}
\begin{figure}
\includegraphics[scale=0.25]{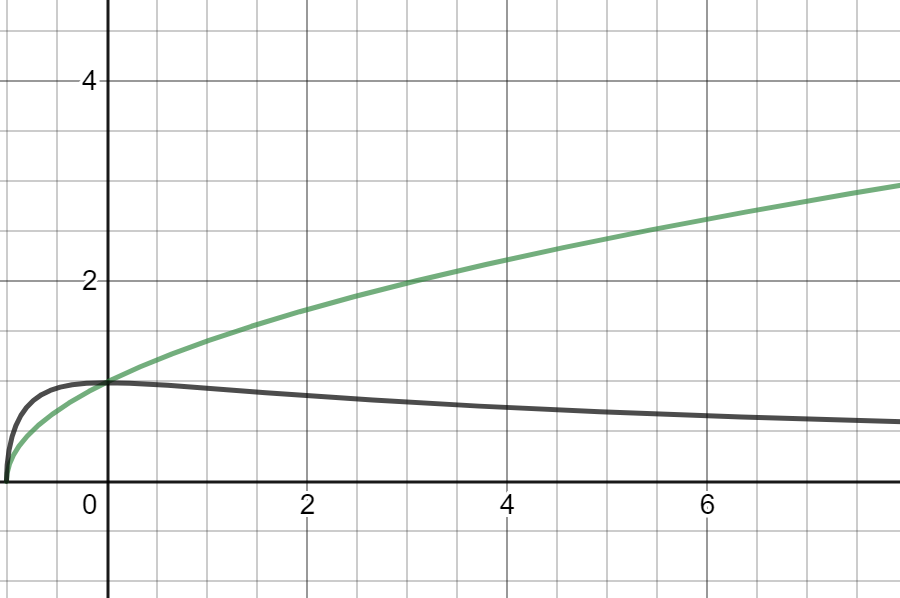}
\caption{The graphs of  $f(\alpha)$  and $g(\alpha)$  for $\epsilon=0.01$. $\omega_{\rm{max}}$ is larger than $\omega_{\rm{sl}}$ for $\alpha>0$. $\omega_{\rm{max}}$ and  $\omega_{\rm{sl}}$ deviate from each other as $\alpha$ increases.}
\label{figure}
\end{figure}
\end{center}

In the last decade it was shown that though extremal Kerr black holes cannot be overspun, nearly extremal Kerr black holes can be overspun into naked singularities by a discrete jump by test particles \cite{Jacobson-Sot} and fields \cite{overspin}. Recently Sorce and Wald considered the second order variations which account for backreaction effects, and showed that overspinning is not possible in a complete second order analysis \cite{w2}. In this section we attempt to overspin nearly-extremal Kerr-MOG black holes by test scalar fields. We parametrise a nearly-extremal Kerr-MOG black hole in the form
\begin{equation}
\frac{J\sqrt{1+\alpha}}{\mathcal{M}^2}=\frac{a\sqrt{1+\alpha}}{\mathcal{M}}=1-\epsilon '^2 \label{nex0}
\end{equation}
where $\epsilon' \ll 1$. (\ref{nex0}) implies that 
\begin{equation}
\delta_{\rm{in}}=\mathcal{M}^2-J\sqrt{1+\alpha}=\mathcal{M}^2 \epsilon'^2 \label{nex}
\end{equation}
As in the case of extremal black holes, we send in a test field from infinity and demand that $\delta_{\rm{fin}}<0$ at the end of the interaction, so that the final parameters of the space-time represent a naked singularity.
\begin{equation}
\delta_{\rm{fin}}=(\mathcal{M}+\delta E)^2 -\left( J +\frac{m}{\omega}\delta E \right)\sqrt{1+\alpha}<0 \label{nex1}
\end{equation}
where we have used that $\delta J=(m/\omega )\delta E$. Again we choose $\delta E=\mathcal{M}\epsilon$ for the energy of the incident wave, and impose (\ref{nex}) to simplify (\ref{nex1}). The condition that $\delta_{\rm{fin}}<0$ can be expressed as
\begin{equation}
\mathcal{M}^2 \epsilon'^2 + \mathcal{M}^2 \epsilon^2 + 2 \mathcal{M}^2 \epsilon-\frac{m}{\omega}\mathcal{M}\epsilon \sqrt{1+\alpha}<0 \label{nex2}
\end{equation}
Using (\ref{nex2}), one directly derives the maximum frequency $\omega_{\rm{max}}$ for a scalar field incident on a nearly extremal Kerr-MOG black hole parametrised as (\ref{nex}), that could overspin the black hole into a naked singularity
\begin{equation}
\omega <\omega_{\rm{max}}=\frac{m\sqrt{1+\alpha}}{\mathcal{M}(2+\epsilon+\epsilon')} \label{wmaxnex}
\end{equation}
As we mentioned in the case of extremal black holes, the condition (\ref{wmaxnex}) is not sufficient for overspinning to occur. We should also demand that the frequency of the incoming wave is larger than the limiting frequency for superradiance. If $(\omega_{\rm{sl}}<\omega_{\rm{max}})$, there exists a range of frequencies $(\omega_{\rm{sl}},\omega_{\rm{max}})$ which can be chosen to overspin a nearly-extremal Kerr-MOG black hole. To compare $\omega_{\rm{sl}}$ and $\omega_{\rm{max}}$, one has to express $\omega_{\rm{sl}}$ for a nearly-extremal black hole in terms of the small parameter $\epsilon$. Notice that for the nearly-extremal Kerr-MOG black hole parametrised as (\ref{nex0})
\begin{equation}
r_+=\mathcal{M}+\sqrt{\frac{\mathcal{M}^2}{1+\alpha}-a^2}=\mathcal{M}\left( 1+\epsilon'\sqrt{\frac{2-\epsilon'^2}{1+\alpha}} \right)
\end{equation}
and
\begin{eqnarray}
r_+^2 + a^2 &=& 2\mathcal{M}r_+-\frac{\alpha}{1+\alpha}\mathcal{M}^2 \nonumber \\
&=&2\mathcal{M}^2\left( 1+\epsilon'\sqrt{\frac{2-\epsilon'^2}{1+\alpha}}-\frac{\alpha}{2(1+\alpha)} \right)
\end{eqnarray}
which leads to
\begin{eqnarray}
\omega_{\rm{sl}}&=& \frac{ma}{r_+^2 + a^2 } \nonumber \\
&=& \frac{m(1-\epsilon'^2)}{2\mathcal{M}\left( \sqrt{1+\alpha}+\epsilon' \sqrt{2-\epsilon'^2}-\frac{\alpha}{2\sqrt{1+\alpha}}\right)} \label{wslnex}
\end{eqnarray}
Though it is not quite manifest in equations (\ref{wmaxnex}) and (\ref{wslnex}), the maximum frequency for an incident wave to overspin a Kerr-MOG black hole is actually larger than the limiting frequency for superradiance. To clarify this we have set $\omega_{\rm{max}}=(m/2\mathcal{M})f(\alpha)$ and $\omega_{\rm{sl}}=(m/2\mathcal{M})g(\alpha)$ and plotted $f(\alpha)$ and $g(\alpha)$ for $\epsilon=\epsilon'=0.01$, in the figure (\ref{figure}). The frequencies $\omega_{\rm{max}}$ and  $\omega_{\rm{sl}}$ almost coincide for $\alpha=0$. However, as $\alpha$ increases the range of frequencies that can be used to overspin a Kerr-MOG black hole enlarges.

\section{Backreaction effects}
In a seminal paper, Will has argued that when test particles approach the event horizon of the black holes, the angular velocity of the event horizon increases due to the dragging of inertial frames \cite{will}. The change in the angular velocity is estimated to be 
\begin{equation}
\Delta \omega=\frac{\delta J}{4M^3}
\label{deltaomega}
\end{equation}
where $\delta J$ is the angular momentum of the test particle or field, and $M$ is the mass of the black hole. We should note that, the black hole itself does not acquire angular momentum before the absorption of the test particle or field, in this process. If this were the case a nearly extremal black hole would be overspun before the absorption of the test particle or field, as the angular momentum parameter increases while the mass is kept constant. However  only the angular velocity of the event horizon increases before the absorption. In particular, one can have a black hole with zero angular momentum with an event horizon with angular velocity given by (\ref{deltaomega}), as stated by Will.

The change in the angular velocity leads to a backreaction  in the scattering problems. Since  the limiting frequency for superradiance increases, the absorption of modes that could lead to the over-spinning of the black hole can be prevented. The backreaction effects in this form was analysed by Hod in \cite{q3}, who argued that the violation of Wccc due to the the tunnelling of scalar particles derived in a previous work \cite{q1} could be prevented as the superradiant limit increases. In this section we calculate the backreaction effects for scalar test fields scattering off Kerr-Mog black holes. We start with the extremal case.

\subsection{Backreaction effects for extremal black holes}
In section (\ref{overspin}) we envisaged an extremal black hole interacting with a test field which carries energy $\delta E=M\epsilon$, and angular momentum $\delta J=(m/\omega)\delta E$. We derived that there exists a range of frequencies $\omega_{\rm{sl}} < \omega < \omega_{\rm{max}}$ that lead to the overspinning of the black hole, where 
\[
\omega_{\rm{max}}=\frac{m\sqrt{1+\alpha}}{\mathcal{M}(2+\epsilon)} 
\]
and
\[
\omega_{\rm{sl}}=\frac{m\sqrt{1+\alpha}}{\mathcal{M}(2+\alpha)} 
\]
To calculate the backreaction effects, let us consider an extremal black hole interacting with a test field with frequency $\omega$ that is arbitrarily close to but slightly less then $\omega_{\rm{sl}}$. If $\alpha$ is larger than $\epsilon$ for the black hole, the test field will be absorbed by the black hole since $\omega>\omega_{\rm{sl}}$, and it will be overspin the black hole into  a naked singularity. However as the test field approaches the black hole the angular velocity of the event horizon will increase by an amount
\begin{equation}
\Delta \omega= \frac{\delta J}{4\mathcal{M}^3}=\frac{\epsilon(2+\epsilon)}{4\mathcal{M} \sqrt{1+\alpha}}
\end{equation}
where we have used that $\omega \simeq \omega_{\rm{sl}}$, and $\delta J=(m/\omega) \delta E$. The limiting value for superradiance will increase  by an amount $\Delta \omega$. If the modified value of  the superradiance limit exceeds the frequency of the incoming field for $\omega\simeq\omega_{\rm{max}}$, it  will exceed the incoming frequency even further for  $\omega_{\rm{sl}}<\omega < \omega_{\rm{max}}$, since $\Delta \omega$ will be larger. Therefore it is critical to calculate the backreaction effects for $\omega\simeq\omega_{\rm{max}}$, for a certain $\alpha$. Now we demand that:
\begin{equation}
\omega_{\rm{sl}}+\Delta \omega \geq \omega
\end{equation}
Explicitly we demand that 
\begin{equation}
\frac{m\sqrt{1+\alpha}}{\mathcal{M}(2+\alpha)}+ \frac{\epsilon(2+\epsilon)}{4\mathcal{M}\sqrt{1+\alpha}} \geq \frac{m\sqrt{1+\alpha}}{\mathcal{M}(2+\epsilon)}
\label{demandext}
\end{equation}
For $\epsilon=0.01$ the condition (\ref{demandext}) is equivalent to
\begin{equation}
\alpha \lesssim 0.0299 \sim 3\epsilon
\label{backext}
\end{equation}
We set $m=1$ in in (\ref{demandext}), since these modes have the highest  absorption probability ignoring the modes with $m=0$ which do not contribute to the angular momentum of the black hole \cite{page}. Thus, the backreaction effects prevent the overspinning of extremal Kerr-Mog black holes if $\alpha \leq 0.0299$. It would be appropriate to elucidate the subject with a numerical example. Let us envisage an extremal Kerr-Mog black hole with $\alpha=0.029$ which is less than the critical value derived in (\ref{backext}). For this black hole we find that
\begin{equation}
\omega_{\rm{max}}=\frac{m}{\mathcal{M}}0.50467; \quad \omega_{\rm{sl}}=\frac{m}{\mathcal{M}}0.49995
\end{equation}
Ignoring backreaction effects, this black hole will be overspun into a naked singularity if it interacts with a test field satisfying $\delta= M\epsilon$, and $\omega \simeq \omega_{\rm{max}}$. However, before the absorption of the test field, the angular velocity of the horizon will increase by an amount
\begin{equation}
\Delta \omega=\frac{\delta J}{4M^3}=\frac{1}{\mathcal{M}}0.00495
\label{deltaomegaex}
\end{equation}
Due to the dragging of the inertial frames, the superradiance limit will be modified. For $m=1$:
\begin{equation}
\omega '_{\rm{sl}}=\frac{1}{\mathcal{M}}\left(0.49995+0.00495\right)=\frac{1}{\mathcal{M}}0.50490
\label{omegaprimeex}
\end{equation}
Since the modified value of the superradiance limit is larger than the frequency of the incident field $(\omega \simeq (1/\mathcal{M}) 0.50467)$, the field will not be absorbed by the black hole; thus the overspinning of the extremal black hole will be prevented. If we choose a smaller value for $\omega$ for the frequency of the incident wave in the range $\omega_{\rm{sl}}<\omega < \omega_{\rm{max}}$, $\Delta \omega$ and $\omega '_{\rm{sl}}$ will be larger than the values derived in (\ref{deltaomegaex}) and (\ref{omegaprimeex}), thus $\omega '_{\rm{sl}}$ will exceed the frequency of the incident wave even further.
Therefore we conclude that the backreaction effects prevent the overspinning of the extremal Kerr-Mog black holes for $\alpha \lesssim 0.0299$. 
\subsection{Backreaction effects for nearly-extremal black holes}
One can proceed the same way to calculate the backreaction effects for nearly extremal black holes. The maximum value for the frequency of a test field to overspin a nearly extremal Kerr-Mog black hole, and the superradiance limit was derived in (\ref{wmaxnex}) and (\ref{wslnex}). Again we demand that the modified value of the superradiance limit exceeds, the frequency of the incoming field, so that the overspinning is prevented. For nearly extremal black holes the increase in the superradiance limit is given by
\begin{equation}
\Delta \omega=\frac{\delta J}{4M^3}=\frac{\epsilon\left( 2+\epsilon +\epsilon' \right)}{4 \mathcal{M}\sqrt{1+\alpha}}
\end{equation}
where we use $\delta E=\mathcal{M}\epsilon$, and $\delta J=(m/\omega)\delta E$ for the incoming field. ($\epsilon'$ is used to parametrise the closeness to extremality.) Again we have substituted the critical value $\omega \simeq \omega_{\rm{max}}$, to derive an expression for $\Delta \omega$. As in the case of extremal black holes we demand that the modified value of the superradiance limit exceeds the frequency of the incoming fields for the challenging modes $\omega_{\rm{sl}}<\omega<\omega_{\rm{max}}$. Setting $\epsilon=\epsilon'=0.01$, one can derive that $\omega_{\rm{sl}}+\Delta \omega$ will be larger than $\omega_{\rm{max}}$ if
\begin{equation}
\alpha \lesssim 0.0119
\end{equation}
Therefore the backreaction effects prevent the overspinning of Kerr-Mog black holes for which $\alpha \lesssim 0.0119$. For a numerical example, let us consider a nearly extremal Kerr-Mog black hole with $\alpha=0.011$. For this black hole we find that 
\begin{equation}
\omega_{\rm{sl}}=\frac{m}{\mathcal{M}}=0.49297; \quad \omega_{\rm{max}}=\frac{m}{\mathcal{M}}0.49776
\end{equation}  
Ignoring backreaction effects, this black hole would be overspun into a naked singularity by a test field with frequency $\omega_{\rm{sl}} <\omega <\omega_{\rm{max}}$. However,  the superradiance limit will be modified by an amount
\begin{equation}
\Delta \omega=\frac{1}{\mathcal{M}}0.00502
\end{equation}
For $m=1$, the modified value of the superradiance limit will be $(1/\mathcal{M})0.49799$, which is larger than $\omega_{\rm{max}}$. Therefore a nearly extremal Kerr-Mog black hole parametrised as (\ref{nex0}) and (\ref{nex}) cannot be overspun by a test field, provided that $\alpha \lesssim 0.0119$, if one considers the increase in the angular velocity of the event horizon due to the interaction with the field. We would like to note that the calculations in this section are  valid for $\epsilon'=0.01$. For smaller values of $\epsilon'$ the values derived for $\Delta \omega$ will approach  the corresponding limit for extremal black holes and the backreactions will work for greater values of $\alpha$ approaching the value derived for extremal black holes.

\section{The validity of the third law for Kerr-MOG black holes}
The laws of black hole dynamics which were proposed by Bardeen, Carter, and Hawking  are based on a connection between  thermodynamics and black hole dynamics  \cite{bardeen}. In this manner the area of the event horizon  and the  surface gravity are analogous to the entropy and the temperature, respectively. The identification of the area of the event horizon with entropy entails that it should not be possible to decrease the area of the event horizon, which had been previously proved by Hawking assuming that no naked singularities exist in the outer region~\cite{area}. Accordingly, it should not be possible to drive a black hole to extremality which would be analogous to decreasing the temperature to absolute zero. After a decade Israel proved the third law of black hole dynamics which states a nearly extremal black hole cannot be driven to extremality in any continuous process~\cite{israel1}. An alternative approach by Dadhich and Karayan also justified the validity of the third law. They showed that the range of the allowed energy and angular momentum ratios to drive a Kerr black hole to extremality, pinches off as one gets arbitrarily close to extremality ~\cite{dadhich}. 

Currently, the validity of the third law is justified for Kerr, Kerr-Newman and Reissner-N\"{o}rdstrom black holes. The derivations by Hubeny, Jacobson-Sotiriou, and D\"{u}zta\c{s}-Semiz that nearly-extremal black holes can be overcharged or overspun into naked singularities \cite{hu,Jacobson-Sot,overspin} should not be interpreted as counter-examples to the third law. These  authors confirm that extremal black holes cannot be overcharged/overspun, which implies that nearly extremal black holes are driven beyond extremality by a discrete jump rather than a continuous process. As one gets arbitrarily close to extremality the allowed ranges of energy, angular momentum, and/or charge for the perturbation vanishes in accord with the derivations of Dadhich and Karayan.   

The analysis for the nearly-extremal Kerr-MOG black holes in the previous section can be exploited to test the validity of the third law. Let us consider a Kerr-MOG black hole arbitrarily close to extremality, which corresponds to the case $\epsilon' \to 0$. The maximum value for the frequency of an incoming scalar field to overspin this Kerr-MOG black hole, and the value of the superradiance limit approach their corresponding values for the extremal case as  $\epsilon' \to 0$  
\begin{equation}
\lim_{\epsilon' \to 0} \omega_{\rm{max}}=\lim_{\epsilon' \to 0} \frac{m\sqrt{1+\alpha}}{\mathcal{M}(2+\epsilon+\epsilon')}=\frac{m\sqrt{1+\alpha}}{\mathcal{M}(2+\epsilon)} \label{wmaxlim}
\end{equation}
\begin{equation}
\lim_{\epsilon' \to 0}\omega_{\rm{sl}}=\frac{m}{2\mathcal{M}\left( \sqrt{1+\alpha} -\frac{\alpha}{2\sqrt{1+\alpha}}\right)}=\frac{m\sqrt{1+\alpha}}{\mathcal{M}(2+\alpha)} 
\end{equation}
A Kerr-MOG black hole arbitrarily close to extremality would become extremal if it absorbed  a test field with frequency $\omega=\omega_{\rm{max}}$ given in (\ref{wmaxlim}), while it would be overspun if $\omega<\omega_{\rm{max}}$ as discussed in the previous section. Contrary to the case of the Kerr family of solutions, the interval $(\omega_{\rm{sl}},\omega_{\rm{max}})$ does not pinch off as the black hole becomes arbitrarily close to extremality. Therefore it first appears that Kerr-MOG black holes can be \emph{continuously} driven to extremality by scalar test fields with frequency $\omega_{\rm{max}}$, which is larger than the superradiance limit $\omega_{\rm{sl}}$ even in the $\epsilon' \to 0$ limit. However we can calculate the increase in the superradiance limit as $\epsilon' \to 0$
\begin{equation}
\lim_{\epsilon' \to 0} \Delta \omega=\frac{\epsilon(2+\epsilon)}{4\mathcal{M}(1+\alpha)}
\end{equation}
which is the corresponding value derived for extremal black holes. The calculations for backreactions imply that if $\alpha \lesssim 0.02999 \sim 3\epsilon $ for nearly extremal black holes, they cannot be driven to extremality by test fields, since the modified frequency for the superradiance limit will exceed the frequency of the incoming field which prevents its absorption. 
Therefore the third law of black hole dynamics is valid for Kerr-MOG black holes which are characterised by a  deformation parameter $\alpha \lesssim 0.03$.

\section{Conclusions}
In this work we applied a test of the weak cosmic censorship conjecture in the interaction of Kerr-MOG black holes with test fields. We restricted ourselves to the case of scalar fields the energy-momentum tensor of which obey the weak energy condition. Our analysis also exploits the fact that superradiance occurs for scalar fields scattering off Kerr-MOG black holes \cite{wondrak}. Superradiance is essential in a scattering process as it determines the lower limit for the frequency of a wave to ensure that it is absorbed by the black hole. In the absence of such a limit the modes carrying low energy and relatively high angular momentum can also be absorbed by the black hole which reinforces the overspinning of the black hole. We have first shown that both extremal and nearly-extremal Kerr-MOG black holes can be overspun by test scalar fields with a frequency slightly above the superradiance limit. The range of the allowed frequencies for the incoming field is extended as the modification parameter $\alpha$ increases. Next we employed the backreaction effects based on the the increase in the limiting frequency for superradiance, which was suggested by Will \cite{will}. We showed that the increase in the superradiance limit, prevents the overspinning of extremal black holes for which $\alpha \lesssim 0.03 \sim 3\epsilon$. We derived this relation by setting $\epsilon=0.01$ in the inequality (\ref{demandext}). The inequality (\ref{demandext}) does not allow us to find an analytical solution for $\alpha$ in terms of $\epsilon$.
However, one can numerical verify that the relation $\alpha \lesssim 3\epsilon$
continues to hold for smaller values of $\epsilon$. (A larger value
for $\epsilon$ would disrupt the test field approximation.) The corresponding value for nearly extremal black holes turns out to be lower: $\alpha \lesssim 0.012$. However we noted that, it approaches the critical value derived for extremal black holes as the black hole approaches extremality. 

In a previous work we had found that, --though extremal Kerr black holes can not-- nearly extremal Kerr black holes can be overspun by test fields \cite{overspin}. We would like to note that the derivation of backreaction effects carried out in this work, directly apply to Kerr black holes with $\alpha=0$. Therefore the backreaction effects based on the argument by Will \cite{will}, also reassure the validity of Wccc for Kerr black holes interacting with test fields.

One would also expect the third law of black hole dynamics to hold for Kerr-MOG black holes analogous to the Kerr case. Our analysis for the nearly-extremal Kerr-MOG black holes imply that the allowed range of frequencies for overspinning to occur does not pinch off even in the $\epsilon' \to 0$ limit. Thus, neglecting backreaction effects, it first appears that a nearly extremal Kerr-MOG black hole that is arbitrarily close to extremality can be continuously driven to extremality by absorbing a test field with frequency $\omega_{\rm{max}}$. However the backreaction effects imply that test fields with frequency $\omega=\omega_{\rm{max}}$ will not be absorbed by nearly extremal Kerr-Mog black holes arbitrarily close to extremality, provided that $\alpha \lesssim 0.03$. Therefore the third law of black hole dynamics is also valid for Kerr-Mog black holes with $\alpha \lesssim 0.03$, which includes the Kerr case with $\alpha=0$.

%
% BibTeX users please use
% \bibliographystyle{}
% \bibliography{}

\end{document}